\documentclass[table]{Interspeech2024}

\interspeechcameraready 

\usepackage[framemethod=TikZ]{mdframed}
\usepackage{amsmath,graphicx}
\usepackage{multirow}
\usepackage{xcolor}

\title{LipGER: Visually-Conditioned Generative Error Correction for Robust Automatic Speech Recognition}

\name[]{Sreyan}{Ghosh}
\name[]{Sonal}{Kumar}
\name[]{Ashish}{Seth}
\name[]{Purva}{Chiniya}
\name[]{Utkarsh}{Tyagi}
\name[]{Ramani}{Duraiswami}
\name[]{Dinesh}{Manocha}

\address{
  University of Maryland, College Park, USA}

\email{\{sreyang, sonalkum, aseth125, pchiniya, utkarsht, ramanid, dmanocha\}@umd.edu
}


\definecolor{darkblue}{rgb}{0.0, 0.0, 1.0}

\begin{document}

\maketitle

\begin{abstract}
Visual cues, like lip motion, have been shown to improve the performance of Automatic Speech Recognition (ASR) systems in noisy environments. We propose LipGER (Lip Motion aided Generative Error Correction), a novel framework for leveraging visual cues for noise-robust ASR. Instead of learning the cross-modal correlation between the audio and visual modalities, we make an LLM learn the task of visually-conditioned (generative) ASR error correction. Specifically, we instruct an LLM to predict the transcription from the $N$-best hypotheses generated using ASR beam-search. This is further conditioned on lip motions. This approach addresses key challenges in traditional AVSR learning, such as the lack of large-scale paired datasets and difficulties in adapting to new domains. We experiment on 4 datasets in various settings and show that LipGER improves the Word Error Rate in the range of 1.1\%-49.2\%. We also release LipHyp, a large-scale dataset with hypothesis-transcription pairs that is additionally equipped with lip motion cues to promote further research in this space~\footnote{Code and Data: https://github.com/Sreyan88/LipGER}.

\end{abstract}

\section{Introduction}

Automatic speech recognition (ASR) enables efficient and accurate transcription of spoken languages. However, in real-world scenarios, speech is generally accompanied by noise from several sources, including background sounds, interfering speakers, and reverberation~\cite{li2014overview}. The performance of ASR systems (trained on clean speech) degrades significantly in such scenarios~\cite{krishna2019speech}. Visual lip motion emerges as a natural source for supervision to improve speech understanding in noisy environments due to its tight coupling with audio~\cite{shi2022learning}. This has motivated much work on Audio-Visual Speech Recognition (AVSR), where researchers have proposed novel ways to fuse visual information with speech for noise-robust ASR~\cite{hong2023watch,shi2022learning,10096893}. However, training AVSR systems that are effective and resilient to acoustic variations like accent, dialect, etc., suffers from a lack of diverse large-scale audio-visual datasets. Acquiring such datasets is prohibitively expensive, as they require to be recorded in controlled environments~\cite{shi2022learning}. On the other hand, existing ASR systems benefit from the relative abundance of speech-only data~\cite{radford2023robust,chen2021gigaspeech}. Thus, an effective solution would be to employ visual information to refine transcriptions from an existing powerful ASR model. To the best of our knowledge, no existing system has effectively been able to use this hypothesis.

\begin{figure}
    \centering
    \includegraphics[width=\columnwidth]{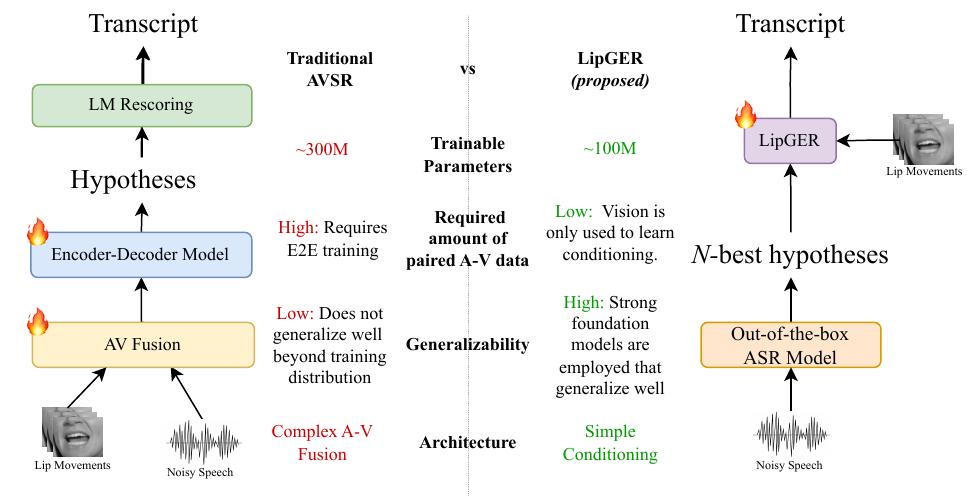}
    \caption{\small Comparison of traditional AVSR methods and LipGER. LipGER benefits over these methods on several aspects by overcoming the requirement of E2E traning.}
    \vspace{-2em}
    \label{fig:cartoon}
\end{figure}
Humans exhibit exceptional resilience to variations in acoustic environments, as our recognition system goes beyond merely processing the received acoustic signals. Typically, we infer unclear or distorted spoken words by leveraging both visual cues and our intrinsic understanding of language. Very recently, Chen~\textit{et al.} \cite{NEURIPS2023_64922674} proposed HyPoradise, a dataset for the task of \textit{ASR generative error correction} with Large Language Models (LLMs). In contrast to the conventional method of using an LM to first re-score all transcriptions in the $N$-best hypothesis list with an LM and then select the best candidate, they propose to predict the transcription directly by prompting an LLM with the $N$-best hypothesis list. The authors argue that their primary motivation stems from the belief that the $N$-best list contains valuable information, as each hypothesis presents a unique textual interpretation of the input speech~\cite{peng2013search,variani2020neural,li2020improving}. Hence, sentences that are disregarded could still contain accurate tokens essential for predicting the correct transcription. Inspired by prior work in uncertainty-estimation-based ASR~\cite{stouten2006model,tran2014fusion} and auto-regressive decoding~\cite{malinin2020uncertainty}, we posit that this hypothesis is more useful in noisy conditions where a higher uncertainty of beam search decoding leads to a more diverse $N$-best hypotheses list that contain accurate tokens~\cite{hu2024large}.

{\noindent \textbf{Main Contributions.}} We propose LipGER, a simple yet effective framework for noise-robust AVSR. As an alternative to the traditional AVSR learning setup, which leverages parallel audio-visual data to jointly learn the correlation between the two modalities, LipGER leverages the visual cues for conditioning an LLM to learn generative error correction for ASR. Similar to Chen~\textit{et al.}, given an $N$-best hypotheses list from noisy source speech, we construct a prompt from it. We then feed this prompt to an LLM to predict the transcription. The LLM is further conditioned on the corresponding lip motion of the speech through multi-modal adapters. We build LipGER on the hypothesis that visual cues play an essential role in language learning in children~\cite{meltzoff1977imitation,davies2009investigating}. Our proposed paradigm of language-space denoising for transcribing noisy speech simplifies noise-robust ASR that usually employs a speech enhancement front-ends. LipGER relies on language-comprehension, that is much easier to learn, benefits from abundant data and can generalize better. Additionally, LipGER employs a much simpler architecture than most existing AVSR systems, which require more complex architectures for learning the audio-visual correlation. Finally, LipGER can be used with out-of-the-box powerful and general-purpose ASR models, thus overcoming the requirement of large-scale paired audio-visual data for different languages, dialects or accents. To summarize, our contributions are as follows:



\begin{enumerate}
    \item We present LipGER, a simple yet effective technique for noise-robust AVSR. Different from traditional AVSR systems, LipGER performs visually-conditioned generative error correction for accurately transcribing noisy speech. 
    
    \item We experiment with LipGER on 4 datasets under various synthetic and real-world noisy conditions and show that LipGER improves WER on these datasets by 1.1\%-49.2\%. 

    \item Additionally, we also open-source LipHyp, a large-scale dataset with hypothesis-transcription pairs equipped with visual lip motion cues to promote further research in this space. 
\end{enumerate}






\section{Related Work}

{\noindent \textbf{Noise Robust and Audio-Visual Speech Recognition.}} Recent noise-robust ASR methods primarily employ speech enhancement front-end to denoise noisy speech before ASR~\cite{fu2019metricgan}. Other methods include domain adversarial training to learn
noise-invariant speech features~\cite{prasad2021investigation} or various pre-processing steps for denoising~\cite{radford2023robust}. On the other hand, it is widely believed that integrating visual cues, like lip motion activity, can help improve the noise-robustness of ASR models~\cite{huang2013audio}. However, they have not been widely adopted due to the lack of large-scale AV datasets recorded in diverse settings with varying languages, accents, etc. Various AVSR systems have been proposed in the literature that aims to improve the interaction between the audio and visual cues to push ASR performance~\cite{10127316}. However, they have not been extensively tested in noisy conditions.
\vspace{0.5mm}

{\noindent \textbf{LM Rescoring and Generative Error Correction (GEC).}} LM rescoring has been widely employed for improving ASR systems. Typically, an external LM is trained separately and utilized to re-score the $N$-best list of hypotheses generated by ASR decoding with beam search. Prior work has explored several ways, including shallow fusion, deliberation, component fusion, and cold fusion. LMs are also widely used for the error correction task in different languages~\cite{wirth2022automatic}, leveraging only the 1-best hypothesis generated by the ASR model~\cite{leng2021fastcorrect}. Furthermore, more recent works utilize a candidates list after decoding for error correction~\cite{mani2020asr,leng2021fastcorrect}. To the best of our knowledge, we are the first to explore multi-modal GEC.

\section{Methodology}
\label{sec:methodology}

\subsection{LipHyp}
\label{subsec:dataset}

In this section, we describe the creation of LipHyp, a dataset with 601k+ unique pairs of hypothesis list $\mathcal{H}_i$ and ground truth transcription $\mathcal{T}_i$ for an audio $\mathcal{A}_i$, where the video $\mathcal{V}_i$ is available for each pair. Specifically, for every $\mathcal{H}_i$ = $\{\mathcal{Y}_0, \cdots, \mathcal{Y}_{N}\}$, we extract the N+1 best hypothesis from the beam search decoder of an ASR system, where $\mathcal{Y}_0$ can be regarded as the ASR prediction for the audio, and the other N hypothesis can be considered as \textit{other hypotheses}. To construct LipHyp, we employ open-source datasets and models described next.

\subsubsection{ASR Models}
We employ 3 state-of-the-art ASR models for $N$-best hypothesis generation, namely WavLM~\cite{chen2022wavlm}, Whisper~\cite{radford2023robust} and Auto-AVSR~\cite{ma2023auto}. All 3 models are unimodal speech-only ASR models, as we did not want additional visual cues to influence our hypotheses list. This is also aligned with our final goal of employing speech-only models for $N$-best hypotheses inference and using only visual cues for post-processing.

{\noindent \textbf{WavLM.}}  We employ the ESPnet toolkit~\cite{watanabe2018espnet} for our WavLM-based ASR system. The architecture consists of two blocks: a self-supervised pre-trained front-end model and the ASR model (433 million parameters in total). The front-end, which is WavLM~\textsubscript{Large} consists of 24
transformer-based encoder layers and is pre-trained using a combination of LibriLight~\cite{librilight} (60k hours of data), Gigaspeech~\cite{chen2021gigaspeech} (10k hours of data), and VoxPopuli~\cite{wang2021voxpopuli} (24k hours of data).
The high-level features from the front-end are fed into the ASR back-end for fine-tuning, which consists of 12 layers of conformer-based~\cite{gulati2020conformer} encoder layers and 6 transformer-based decoder layers. The ASR system is fine-tuned on 960-hours LibriSpeech data.
\vspace{0.5mm}

{\noindent \textbf{Whisper.}} We employ the Whisper~\textsubscript{Base} model to generate hypotheses. The encoder-decoder transformer-based architecture has 74 million parameters and is trained on 680,000 hours of multilingual-weakly labeled speech data from the web.
\vspace{0.5mm}

{\noindent \textbf{Auto-AVSR.}} Similarly, the ASR encoder consists of a 1D ResNet-18~\cite{he2016deep} followed by a Conformer decoder. The model is trained using joint CTC/attention training~\cite{Kim2016JointCB} and trained using LRS2~\cite{son2017lip}, LRS3~\cite{son2017lip}, VoxCeleb2~\cite{chung2018voxceleb2} and AVSpeech~\cite{ephrat2018looking}.

\subsubsection{Datasets}

We employ 3 datasets for training, namely LRS2, LRS3, and Facestar~\cite{yang2022audiovisual}. For testing, in addition to the 3 datasets we also add EasyCom~\cite{donley2021easycom}, a real-world noisy dataset recorded in an egocentric view. To generate the $N$-best hypothesis from these datasets, we simulate
noisy speech for all splits. To achieve this, we first add room reverberation by convolving clean speech with impulse responses from the MIT Impulse Response Survey~\cite{doi:10.1073/pnas.1612524113}. It consists of 270 impulse responses collected from different locations that volunteers encountered in their daily lives and reflect a variety of standard acoustic settings. Subsequently, we add random audio samples from the VoxCeleb2 dataset to simulate interfering speakers and random audio samples from Audioset~\cite{gemmeke2017audio} to simulate background noise. For the background noise, we adjust the signal-to-noise ratio randomly between 0 and 40 dB with respect to the clean signal.





\subsection{LipGER}

\begin{figure*}[t]
    \centering
    \includegraphics[width=\textwidth]{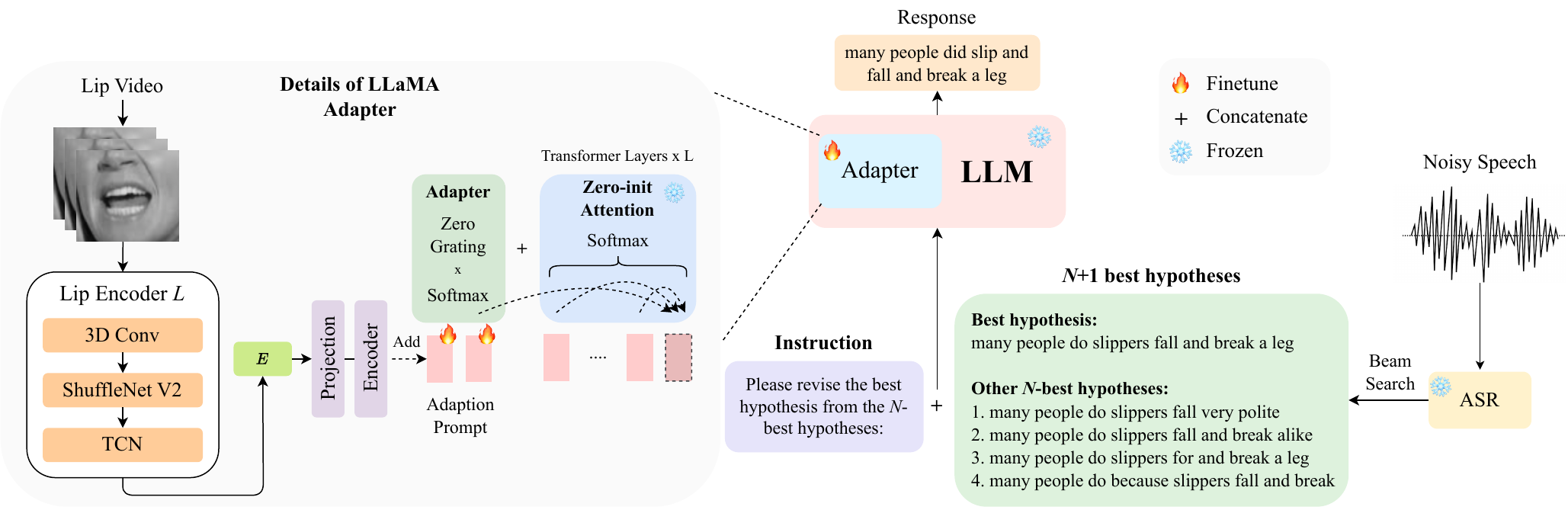}
    \caption{\small Illustration of \textbf{LipGer}. LipGER performs generative error correction on the $N$-best hypotheses generated by beam search decoding from an ASR model. LipGER is built on an LLM, which performs multi-modal reasoning by conditioning on lip motions. Specifically, we build a prompt from the (N+1)-best hypotheses list, which instructs the LLM to rewrite the best hypothesis using the other hypotheses. The lip motion is conditioned on the LLM using multi-modal adapters with encodings obtained from a lip encoder $L$, and only the adapters and $L$ are trained during the fine-tuning stage.}
    \vspace{-1.5em}
    \label{fig:diagram}
\end{figure*}

Fig.~\ref{fig:diagram} illustrates the working of LipGER. First, we input a noisy audio to the ASR system to generate $N+1$-best hypotheses via beam search decoding. Next, we construct an instruction from the hypothesis list and input it to the LLM to generate an output. The LLM generation is conditioned on the corresponding lip motion, which is fed as visual information to the model. In the next sections, we describe our architecture in detail.

\subsubsection{Lip Motion Encoder}
Similar to state-of-the-art lip-reading systems~\cite{martinez2020lipreading}, our lip motion encoder $L$ takes $M$ mouth regions of interest (ROIs) as input. These ROIs are obtained by performing mouth cropping on the input video using facial landmark detection and alignment in a method similar to ~\cite{ma2022visual}. $L$ consists of a 3D convolutional layer followed by a ShuffleNet v2~\cite{ma2018shufflenet} network to extract a time-indexed sequence of feature vectors. They are then processed by a temporal convolutional network (TCN) to extract the final lip motion feature map $E \in \mathbb{R}^{V\times M}$, where $V$ is the fixed sequence length of the feature. $E$ is then fed to the LLM.

\subsubsection{Lip Motion Conditioned LLM}

{\noindent \textbf{Large Language Model (LLM).}} We employ an LLM for the task of GER due to its language generation and reasoning capabilities~\cite{zhao2023survey}. We only employ the pre-trained version of an LLM (similar to the general practice of employing a pre-trained LM to rescore the hypotheses) and fine-tune it using our LipHyp dataset to complete the GER task. We build instruction-response pairs from instances in the LipHyp dataset for training and evaluation. Once the model is trained, the evaluation phase is similar to prompting, where we prompt the LLM for the transcription. We next describe how we build training instances from LipHyp for fine-tuning.
\vspace{0.5mm}

\begin{table*}[!t]
\caption{\small Result comparison of LipGER with other baselines. \textbf{Left:} Comparison of LipGER with text-only GER, LM-based restoring (+LM$_{score}$) and speech enhancement front-end (+Enhance). ``$LR$'' denotes LipGER trained on the low-resource split of the same dataset as the test set. ``$All$'' denotes LipGER trained on the entire LipHyp dataset. LipGER improves over baselines in the range of 1.1\%-49.2\%. \textbf{Right:} Comparison of LipGER with other AVSR systems. LipGER improves over baselines in the range of 1.9\%-18.4\% , is more robust to domain shifts and significantly outperforms them in low-resource settings.}
\label{table:results_llama2}
\vspace{-0.5em}
\centering
\resizebox{1.4\columnwidth}{!}{
\begin{tabular}{cc|c|c|c|c|c|ccc}
\toprule[1.2pt]
\multicolumn{2}{c|}{\multirow{2}{*}{Test Set}} & \multirow{1}{*}{Model} & \multirow{1}{*}{Baseline} & \multirow{2}{*}{+LM$_{score}$} & \multirow{2}{*}{+Enhance} & \multirow{2}{*}{+GER} & + LipGER$_{LR}$ & LipGER & LipGER$_{All}$ \\
& & & & & & & \textit{(ours)}& \textit{(ours)} & \textit{(ours)}\\ 
\midrule[1.2pt]
\multirow{6}{*}{LRS2} & \multirow{3}{*}{\emph{test-real}} & Whisper & $33.4$ & $32.8_{\textcolor{darkblue}{-1.8\%}}$ & $32.9_{\textcolor{darkblue}{-1.5\%}}$ & $29.7_{\textcolor{darkblue}{-11.8\%}}$ & \cellcolor{gray!10} ${26.3_{\textcolor{darkblue}{-21.26\%}}}$ & \cellcolor{gray!10}$\underline{25.9_{\textcolor{darkblue}{-22.5\%}}}$ & \cellcolor{gray!10} $\bm{25.1_{\textcolor{darkblue}{-24.8\%}}}$\\     
& & WavLM & $5.9$  & $5.6_{\textcolor{darkblue}{-5.1\%}}$ & $6.2_{\textcolor{darkblue}{+5.1\%}}$ & $7.0_{\textcolor{darkblue}{+18.6\%}}$ & \cellcolor{gray!10}$\underline{5.4_{\textcolor{darkblue}{-8.5\%}}}$ & \cellcolor{gray!10} $\underline{5.4_{\textcolor{darkblue}{-8.5\%}}}$ & \cellcolor{gray!10}$\bm{5.3_{\textcolor{darkblue}{-10.2\%}}}$ \\
& & Auto-AVSR & $\underline{1.5}$  & $\underline{1.5_{\textcolor{darkblue}{-0.0\%}}}$ & $1.9_{\textcolor{darkblue}{+26.7\%}}$ & $\underline{1.5_{\textcolor{darkblue}{-0.0\%}}}$ & \cellcolor{gray!10}$\underline{1.5_{\textcolor{darkblue}{-0.0\%}}}$ & \cellcolor{gray!10}$\bm{1.4_{\textcolor{darkblue}{-6,.7\%}}}$ & \cellcolor{gray!10}$\bm{1.4_{\textcolor{darkblue}{-6.7\%}}}$ \\ \cmidrule{2-10}
& \multirow{3}{*}{\emph{test-simu}} & Whisper & $74.2$ & $72.6_{\textcolor{darkblue}{-2.1\%}}$ & $67.1_{\textcolor{darkblue}{-9.6\%}}$ & $48.2_{\textcolor{darkblue}{-35.0\%}}$ & \cellcolor{gray!10}$43.7_{\textcolor{darkblue}{-41.1\%}}$ & \cellcolor{gray!10}$\underline{42.8_{\textcolor{darkblue}{-42.3\%}}}$ & \cellcolor{gray!10}$\bm{40.1_{\textcolor{darkblue}{-46.0\%}}}$ \\
& & WavLM & $19.1$  & $17.9_{\textcolor{darkblue}{-6.3\%}}$ & $18.4_{\textcolor{darkblue}{-3.7\%}}$ & $21.7_{\textcolor{darkblue}{13.6\%}}$ & \cellcolor{gray!10}$17.9_{\textcolor{darkblue}{-6.3\%}}$ & \cellcolor{gray!10}$\underline{17.0_{\textcolor{darkblue}{-11.0\%}}}$ & \cellcolor{gray!10}$\bm{16.8_{\textcolor{darkblue}{-12.04\%}}}$ \\
& & Auto-AVSR & $26.4$  & $25.8_{\textcolor{darkblue}{-2.3\%}}$ &  $26.2_{\textcolor{darkblue}{-0.8\%}}$ & $24.1_{\textcolor{darkblue}{-8.71\%}}$ & \cellcolor{gray!10}$21.8_{\textcolor{darkblue}{-17.42\%}}$ & \cellcolor{gray!10}$\underline{21.1_{\textcolor{darkblue}{-20.0\%}}}$ & \cellcolor{gray!10}$\bm{20.2_{\textcolor{darkblue}{-23.5\%}}}$ \\
\midrule[0.75pt]
\multirow{6}{*}{LRS3} & \multirow{3}{*}{\emph{test-real}} & Whisper & $19.0$ & $18.2_{\textcolor{darkblue}{-4.2\%}}$ &  $18.8_{\textcolor{darkblue}{-1.1\%}}$ &  $15.5_{\textcolor{darkblue}{-18.4\%}}$ & \cellcolor{gray!10}${14.8_{\textcolor{darkblue}{-22.1\%}}}$ & \cellcolor{gray!10}$\underline{14.6_{\textcolor{darkblue}{-23.2\%}}}$ & \cellcolor{gray!10}$\bm{14.4_{\textcolor{darkblue}{-24.2\%}}}$\\
& & WavLM & $4.4$ & $4.4_{\textcolor{darkblue}{-0.0\%}}$ &  $4.2_{\textcolor{darkblue}{-4.5\%}}$ & $4.1_{\textcolor{darkblue}{-6.8\%}}$ & \cellcolor{gray!10}$\underline{4.0_{\textcolor{darkblue}{-9.0\%}}}$ &\cellcolor{gray!10}$\underline{4.0_{\textcolor{darkblue}{-9.0\%}}}$ & \cellcolor{gray!10}$\bm{3.9_{\textcolor{darkblue}{-11.4\%}}}$\\
& & Auto-AVSR & $1.2$ & $1.1_{\textcolor{darkblue}{-8.3\%}}$ &  $1.3_{\textcolor{darkblue}{+8.3\%}}$ & $\bm{1.1_{\textcolor{darkblue}{-8.3\%}}}$ & \cellcolor{gray!10}$\bm{1.1_{\textcolor{darkblue}{-8.3\%}}}$ & \cellcolor{gray!10}$\bm{1.1_{\textcolor{darkblue}{-8.3\%}}}$ & \cellcolor{gray!10}$\bm{1.1_{\textcolor{darkblue}{-8.3\%}}}$\\ \cmidrule{2-10}
& \multirow{3}{*}{\emph{test-simu}} & Whisper & $30.9$ & $28.3_{\textcolor{darkblue}{-8.4\%}}$ & $30.2_{\textcolor{darkblue}{-2.3\%}}$ & $20.2_{\textcolor{darkblue}{-34.6\%}}$ & \cellcolor{gray!10}${17.3_{\textcolor{darkblue}{-44.0\%}}}$ & \cellcolor{gray!10}$\underline{16.1_{\textcolor{darkblue}{-47.9\%}}}$ & \cellcolor{gray!10}$\bm{15.7_{\textcolor{darkblue}{-49.2\%}}}$\\
& & WavLM & $17.7$ & $17.5_{\textcolor{darkblue}{-1.1\%}}$ & $17.0_{\textcolor{darkblue}{-4.0\%}}$ & $18.5_{\textcolor{darkblue}{+4.5\%}}$ & \cellcolor{gray!10}$\underline{14.7_{\textcolor{darkblue}{-16.9\%}}}$ & \cellcolor{gray!10}$\bm{14.6_{\textcolor{darkblue}{-17.5\%}}}$ & \cellcolor{gray!10}$\underline{14.7_{\textcolor{darkblue}{-16.9\%}}}$\\
& & Auto-AVSR & $16.9$ & $16.0_{\textcolor{darkblue}{-5.3\%}}$ & $15.5_{\textcolor{darkblue}{-8.3\%}}$ & $16.0_{\textcolor{darkblue}{-5.33\%}}$ &\cellcolor{gray!10} ${12.9_{\textcolor{darkblue}{-23.7\%}}}$ & \cellcolor{gray!10}$\underline{12.2_{\textcolor{darkblue}{-27.8\%}}}$ & \cellcolor{gray!10}$\bm{11.7_{\textcolor{darkblue}{-30.8\%}}}$ \\
\midrule[0.75pt]
\multirow{6}{*}{Facestar} & \multirow{3}{*}{\emph{test-real}} & Whisper & $17.2$ & $16.8_{\textcolor{darkblue}{-2.3\%}}$ & $16.1_{\textcolor{darkblue}{-6.4\%}}$ & $12.9_{\textcolor{darkblue}{-25.0\%}}$ & \cellcolor{gray!10}$\bm{12.4_{\textcolor{darkblue}{-27.9\%}}}$ & \cellcolor{gray!10}$\underline{12.5_{\textcolor{darkblue}{-27.3\%}}}$ & \cellcolor{gray!10}$\underline{12.5_{\textcolor{darkblue}{-27.3\%}}}$\\ 
& & WavLM & $25.7$ & $24.0_{\textcolor{darkblue}{-6.6\%}}$ & $23.9_{\textcolor{darkblue}{-7.0\%}}$ & $18.2_{\textcolor{darkblue}{-29.2\%}}$ & \cellcolor{gray!10}${16.8_{\textcolor{darkblue}{-34.6\%}}}$ & \cellcolor{gray!10}$\underline{16.5_{\textcolor{darkblue}{-35.8\%}}}$ & \cellcolor{gray!10}$\bm{15.5_{\textcolor{darkblue}{-36.75\%}}}$\\
& & Auto-AVSR & $20.5$ & $19.6_{\textcolor{darkblue}{-4.4\%}}$ & $18.2_{\textcolor{darkblue}{-11.2\%}}$ & $18.1_{\textcolor{darkblue}{-11.7\%}}$ & \cellcolor{gray!10}$\bm{14.0_{\textcolor{darkblue}{-31.7\%}}}$ & \cellcolor{gray!10}$\underline{14.2_{\textcolor{darkblue}{-30.7\%}}}$ & \cellcolor{gray!10}$\underline{14.2_{\textcolor{darkblue}{-30.7\%}}}$ \\ \cmidrule{2-10}
& \multirow{3}{*}{\emph{test-simu}} & Whisper & $57.5$ & $55.5_{\textcolor{darkblue}{-3.5\%}}$ & $53.6_{\textcolor{darkblue}{-6.8\%}}$ & $55.2_{\textcolor{darkblue}{-4.0\%}}$ & \cellcolor{gray!10}$\underline{50.5_{\textcolor{darkblue}{-12.2\%}}}$ & \cellcolor{gray!10}$\underline{50.5_{\textcolor{darkblue}{-12.2\%}}}$ & \cellcolor{gray!10}$\bm{46.0_{\textcolor{darkblue}{-20.0\%}}}$ \\
& & WavLM & $53.7$ & $51.1_{\textcolor{darkblue}{-4.8\%}}$ & $51.0_{\textcolor{darkblue}{-5.0\%}}$ & $50.9_{\textcolor{darkblue}{-5.2\%}}$ & \cellcolor{gray!10}${44.0_{\textcolor{darkblue}{-18.1\%}}}$ &\cellcolor{gray!10} $\underline{43.2_{\textcolor{darkblue}{-19.6\%}}}$ & \cellcolor{gray!10}$\bm{40.8_{\textcolor{darkblue}{-24.0\%}}}$ \\
& & Auto-AVSR & $50.1$ & $49.0_{\textcolor{darkblue}{-2.2\%}}$ & $48.7_{\textcolor{darkblue}{-2.8\%}}$ & $48.4_{\textcolor{darkblue}{-3.4\%}}$ & \cellcolor{gray!10}${43.1_{\textcolor{darkblue}{-14.0\%}}}$ & \cellcolor{gray!10}$\underline{41.5_{\textcolor{darkblue}{-17.2\%}}}$ & \cellcolor{gray!10}$\bm{39.0_{\textcolor{darkblue}{-22.2\%}}}$\\
\midrule
\multirow{3}{*}{EasyCom} & \multirow{3}{*}{\emph{test-real}} & Whisper & $102.3$  & $99.4_{\textcolor{darkblue}{-2.8\%}}$ & $100.3_{\textcolor{darkblue}{-2.0\%}}$ & ${85.8_{\textcolor{darkblue}{-16.1\%}}}$ & \cellcolor{gray!10}$\underline{76.7_{\textcolor{darkblue}{-25.0\%}}}$ & \cellcolor{gray!10}$\bm{76.6_{\textcolor{darkblue}{-25.1\%}}}$ & \cellcolor{gray!10} $\bm{76.6_{\textcolor{darkblue}{-25.1\%}}}$\\ 
& & WavLM & $102.1$  & $99.9_{\textcolor{darkblue}{-2.2\%}}$ & $100.9_{\textcolor{darkblue}{-1.2\%}}$ & $94.7_{\textcolor{darkblue}{-7.2\%}}$ & \cellcolor{gray!10}${94.5_{\textcolor{darkblue}{-7.4\%}}}$ & \cellcolor{gray!10}$\underline{94.3_{\textcolor{darkblue}{-7.6\%}}}$ & \cellcolor{gray!10}$\bm{94.1_{\textcolor{darkblue}{-7.8\%}}}$\\
& & Auto-AVSR & $79.6$  & $76.0_{\textcolor{darkblue}{-4.5\%}}$ & $80.4_{\textcolor{darkblue}{+1.0\%}}$ & $68.0_{\textcolor{darkblue}{-14.6\%}}$ & \cellcolor{gray!10}$\underline{63.9_{\textcolor{darkblue}{-19.7\%}}}$ & \cellcolor{gray!10}$\underline{63.9_{\textcolor{darkblue}{-19.7\%}}}$ & \cellcolor{gray!10}$\bm{62.9_{\textcolor{darkblue}{-21.0\%}}}$\\
\bottomrule[1.2pt]
\end{tabular}}
\hspace{0.15em}
\resizebox{0.7\columnwidth}{!}{
\begin{tabular}{c|cccc}
\toprule[1.2pt] 
Dataset & LRS2 & LRS3 & Facestar & LRS3 $\rightarrow$ EasyCom \\ \midrule[1.2pt]
AV-HuBERT & 28.1 & 20.6 & 48.7 & 94.3 \\
AV-HuBERT$_{LR}$ & 30.2 & 21.7 & 51.6 & 97.6\\ \hline
Auto-AVSR & 23.0 & $\bm{10.3}$ & 47.8 & 82.3 \\
Auto-AVSR$_{LR}$ & 24.7 & 16.1 & 56.1 & 101.5 \\ \hline
TM-CTC & 29.9 & 19.4 & 58.4 & 93.1\\
TM-CTC$_{LR}$ & 35.8 & 24.0 & 62.3 & 98.5\\ \hline
MOCO+Wav2Vec2 & 27.3 & 17.5 & 54.0 & 85.8 \\
MOCO+Wav2Vec2$_{LR}$ & 34.0 & 26.4 & 58.7 & 89.6 \\ \hline
Xu \textit{et al.} & 30.7 & 25.9 & 56.1 & 101.8 \\
Xu \textit{et al.}$_{LR}$ & 38.4 & 33.4 & 65.4 & 103.8 \\ \hline
\cellcolor{gray!10} LipGer \textit{(ours)} &\cellcolor{gray!10} $\bm{21.1}$ &\cellcolor{gray!10} $\underline{12.2}$ &\cellcolor{gray!10} $\bm{41.5}$ & \cellcolor{gray!10}$\bm{63.9}$\\
\cellcolor{gray!10} LipGer$_{LR}$ \textit{(ours)} & \cellcolor{gray!10}$\underline{21.8}$&\cellcolor{gray!10} 12.9 & \cellcolor{gray!10} $\underline{43.1}$ &\cellcolor{gray!10} $\underline{66.7}$\\
\bottomrule[1.2pt]
\end{tabular}}
\vspace{-0.25cm}
\end{table*}

{\noindent \textbf{Instruction Construction.}} We resort to the following template for instruction construction:

\begin{mdframed}[backgroundcolor=gray!20, linecolor=white, innerleftmargin=10pt, innerrightmargin=10pt, innertopmargin=6pt, innerbottommargin=6pt]

Below is the best-hypotheses transcribed from a speech recognition system. Please try to revise it using the words that are only included in the other-hypothesis, and write the response for the true transcription.

{\noindent \textbf{Best-hypothesis:}} you a very kind day

{\noindent \textbf{Other-hypotheses:}} you are very kind day, you have very kind day, $\cdots$ $N$ hypotheses 

{\noindent \textbf{Response:}} you are very kind

\end{mdframed}

\vspace{0.5mm}

{\noindent \textbf{Multi-modal Adapters.}} As described earlier, LipGER completes the task of GER on noisy speech with multi-modal reasoning on lip motions. We design our architecture such that the fine-tuning can be performed in a parameter-efficient way, which benefits over existing AVSR systems that require a model to be trained from scratch for taking visual cues into consideration. Several parameter-efficient fine-tuning techniques have been proposed in literature~\cite{xu2023parameter}, and we resort to adapter-based fine-tuning inspired by Zhang \textit{et al.}~\cite{zhang2024llamaadapter}. Precisely, for every layer $l$ of the $L$ transformer layers, we first construct two learnable prompts, $P^v_l \in \mathbb{R}^{K\times C}$ and $P^a_l \in \mathbb{R}^{K\times C}$, where $K$ is the prompt length and $C$ is the feature dimension. Then we project the lip motion encoding $E$ into the same dimension as $P^v_l$  and concat them together along the sequence dimension as follows:

\begin{equation}
I_l=\operatorname{Concat}(\operatorname{Projection}(E),P^v_l)
\end{equation}

where $I_l \in \mathbb{R}^{(K+V)\times C}$. Next, we pass $I_l$ through a transformer encoder with $T$ layers and slice $I_l$ back to a sequence length of $K$ as follows:

\begin{equation}
    G_l=\operatorname{Encoder}(I_l)[:K, :]
\end{equation}

Finally, we add $G_l$ and $P^a_l$ to obtain the final adaptation prompt $A_l \in \mathbb{R}^{K\times C}$. $A_l$ is then concatenated as a prefix to the I-length word tokens $T_l$ as follows:

\begin{equation}
    \left[A_l ; T_l\right] \in \mathbb{R}^{(K+I) \times C}
\end{equation}

We fine-tune the LLM using standard cross-entropy loss and keep only the adapters, the lip motion encoder, and the visual encoder trainable ($\approx$200M parameters).

\section{Experiments and Results}
\subsection{Experimental Setup}

{\noindent \textbf{Hyper-parameters.}} We experiment with the pre-trained only versions of TinyLlaMa~\cite{zhang2024tinyllama} (1.1B parameters) and Phi-1.5~\cite{textbooks2} (1.3B parameters) for LLMs. We do not experiment with larger LLMs due to resource constraints. We fine-tune for 2 epochs with a constant learning rate of 5e-3, a weight decay of 0.02, and a batch size of 32. For adapters, we employ $K$=15, and our vision encoder has 4 transformer encoder layers.
\vspace{0.5mm}

{\noindent \textbf{Baselines.}} We first compare LipGER with other rescoring methods, namely (i) GER~\cite{NEURIPS2023_64922674} -- LipGER without visual cues (ii) LM$_{rank}$ -- We use the same LLM as GER for rescoring the $N$-best hypotheses and finally take the hypothesis with the best score averaged across the LLM and ASR model scores. In an alternative setup for the noisy speech, we also employ a speech enhancement front-end, a HiFi-GAN~\cite{su2020hifi}, to denoise the noisy speech before passing it to the ASR model. For comparison with other AVSR methods, we use Auto-AVSR, AV-HuBERT~\cite{shi2022learning}, TM-CTC~\cite{afouras2018deep}, MOCO+Wav2Vec2~\cite{pan-etal-2022-leveraging} and Xu \textit{et al.}~\cite{xu2020discriminative}. \textbf{\textit{We employ recent baselines with open-source code available as we re-trained all models for evaluation.}}
\vspace{0.5mm}

{\noindent \textbf{Dataset Settings.}} For training our models, we use either the entire dataset or a low-resource setting as follows: LRS3 (407 hours/30 hours), LRS2 (195 hours/29 hours), Facestar (10 hours/2 hours). For evaluation, we test on both the original (\textit{test-real}) and the simulated noisy test sets (\textit{test-simu}) for all the above datasets together with EasyCom. All methods have been evaluated on the standard Word Error Rate (WER) metric.

\subsection{Results}
Table~\ref{table:results_llama2} (left) compares LipGER with 3 baselines, namely, text-only GER, LM-based rescoring (+LM$_{score}$) and speech enhancement front-end (+Enhance). We report scores for 3 versions of LipGER -- LipGER$_{LR}$ is trained on low-resource train splits of the same dataset the Test Set is taken from, LipGER is trained on the entire train split of the of the same dataset the Test Set is taken from and LipGER$_{All}$ is trained on the entire LipHyp (described in Section~\ref{subsec:dataset}). For EasyCom, we employ LipGER and LipGER$_{LR}$ trained on LRS3. All (speech-only) ASR models employed are pre-trained (training settings are described in Section~\ref{subsec:dataset}) and we don't fine-tune them on the downstream dataset. LipGER outperforms all our baselines by 6.3\%-44.0\%. LipGER$_{All}$ improves over LipGER by 0.4\%-1.3\%. LipGER$_{LR}$ also outperforms all our baselines with only $\approx$1.5\% drop in WER over LipGER, thereby proving its efficacy in low-resource scenarios with limited lip-motion cues available for learning AVSR. On an average, test-real splits have lower WER performance than test-sim splits with EasyCom having the highest WER due to significant motion blur and distant speakers.


Table~\ref{table:results_llama2} (right) compares LipGER with other AVSR approaches. LipGER outperforms all baselines by 1.9\%-18.4\%, showing the most significant improvements on the EasyCom dataset. This dataset, unlike LRS3, features speakers at a distance recorded from a first-person perspective, leading to a notable difference in the evaluation domain. Despite this, LipGER exhibits superior resilience to such domain shifts compared to other AVSR models. This enhanced robustness is attributed to LipGER's reliance on adapting to language-space rather than speech-space, which inherently offers greater stability against domain variations.



\section{Conclusion and Limitations}
\begin{figure}[t]
    \centering
    \includegraphics[width=\columnwidth]{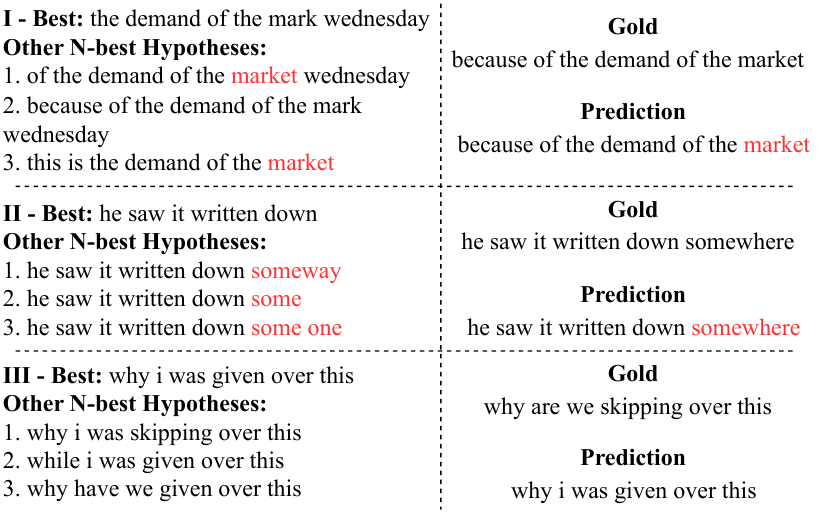}
    \caption{\small Illustration of predictions by LipGER. Instances I. and II. show positive cases where LipGER predicts the correct transcription by revising the best hypothesis with accurate tokens hidden in the other $N$-best hypotheses. Case III. shows a failure case highlighting the limitation of LipGER, which fails to generate the correct transcription due to a lack of sufficient contextual cues in the other $N$-best hypotheses.}
    \vspace{-1.75em}
    \label{fig:qual}
\end{figure}

In this paper, we present LipGER, a novel system for learning robust audio-visual speech recognition. In contrast to prior methods, which employ visual cues for learning speech recognition, LipGER leverages visual cues for learning ASR Generative Error Correction. To achieve this, LipGER employs an LLM conditioned on lip motions using multi-modal adapters. LipGER is simple, requires minimal training parameters, and overcomes several limitations in the current audio-visual ASR learning paradigm. We show that LipGER outperforms most baselines from the literature on noise robust speech recognition by significant margins in low and high-resource settings.

Some limitations of LipGER include: (1) the requirement of an LLM for the pipeline and (2) the predictions might encode biases learned by the LLM during pre-training. As part of future work, we would like to find ways to overcome these limitations.


\bibliographystyle{IEEEtran}
\bibliography{mybib}

\begin{thebibliography}{10}
\providecommand{\url}[1]{#1}
\csname url@samestyle\endcsname
\providecommand{\newblock}{\relax}
\providecommand{\bibinfo}[2]{#2}
\providecommand{\BIBentrySTDinterwordspacing}{\spaceskip=0pt\relax}
\providecommand{\BIBentryALTinterwordstretchfactor}{4}
\providecommand{\BIBentryALTinterwordspacing}{\spaceskip=\fontdimen2\font plus
\BIBentryALTinterwordstretchfactor\fontdimen3\font minus \fontdimen4\font\relax}
\providecommand{\BIBforeignlanguage}[2]{{%
\expandafter\ifx\csname l@#1\endcsname\relax
\typeout{** WARNING: IEEEtran.bst: No hyphenation pattern has been}%
\typeout{** loaded for the language `#1'. Using the pattern for}%
\typeout{** the default language instead.}%
\else
\language=\csname l@#1\endcsname
\fi
#2}}
\providecommand{\BIBdecl}{\relax}
\BIBdecl

\bibitem{li2014overview}
J.~Li, L.~Deng, Y.~Gong, and R.~Haeb-Umbach, ``An overview of noise-robust automatic speech recognition,'' \emph{IEEE/ACM Transactions on Audio, Speech, and Language Processing}, vol.~22, no.~4, pp. 745--777, 2014.

\bibitem{krishna2019speech}
G.~Krishna, C.~Tran, J.~Yu, and A.~H. Tewfik, ``Speech recognition with no speech or with noisy speech,'' in \emph{IEEE ICASSP 2019}.

\bibitem{shi2022learning}
B.~Shi, W.-N. Hsu, K.~Lakhotia, and A.~Mohamed, ``Learning audio-visual speech representation by masked multimodal cluster prediction,'' \emph{arXiv preprint arXiv:2201.02184}, 2022.

\bibitem{hong2023watch}
J.~Hong \emph{et~al.}, ``Watch or listen: Robust audio-visual speech recognition with visual corruption modeling and reliability scoring,'' \emph{arXiv preprint arXiv:2303.08536}, 2023.

\bibitem{10096893}
J.~Li \emph{et~al.}, ``Robust audio-visual asr with unified cross-modal attention,'' in \emph{IEEE ICASSP 2023}.

\bibitem{radford2023robust}
A.~o. Radford, ``Robust speech recognition via large-scale weak supervision,'' in \emph{ICML 2023}.

\bibitem{chen2021gigaspeech}
G.~Chen \emph{et~al.}, ``Gigaspeech: An evolving, multi-domain asr corpus with 10,000 hours of transcribed audio,'' \emph{arXiv preprint arXiv:2106.06909}, 2021.

\bibitem{NEURIPS2023_64922674}
C.~CHEN \emph{et~al.}, ``Hyporadise: An open baseline for generative speech recognition with large language models,'' in \emph{NeruIPS 2023}.

\bibitem{peng2013search}
F.~Peng \emph{et~al.}, ``Search results based n-best hypothesis rescoring with maximum entropy classification,'' in \emph{IEEE ASRU 2013}.

\bibitem{variani2020neural}
E.~Variani \emph{et~al.}, ``Neural oracle search on n-best hypotheses,'' in \emph{IEEE ICASSP 2020}.

\bibitem{li2020improving}
M.~Li \emph{et~al.}, ``Improving spoken language understanding by exploiting asr n-best hypotheses,'' \emph{arXiv preprint arXiv:2001.05284}, 2020.

\bibitem{stouten2006model}
V.~Stouten, P.~Wambacq \emph{et~al.}, ``Model-based feature enhancement with uncertainty decoding for noise robust asr,'' \emph{Speech communication}, vol.~48, no.~11, pp. 1502--1514, 2006.

\bibitem{tran2014fusion}
D.~T. Tran, E.~Vincent, and D.~Jouvet, ``Fusion of multiple uncertainty estimators and propagators for noise robust asr,'' in \emph{IEEE ICASSP 2014}.

\bibitem{malinin2020uncertainty}
A.~Malinin \emph{et~al.}, ``Uncertainty estimation in autoregressive structured prediction,'' \emph{arXiv preprint arXiv:2002.07650}, 2020.

\bibitem{hu2024large}
Y.~Hu \emph{et~al.}, ``Large language models are efficient learners of noise-robust speech recognition,'' in \emph{ICLR 2024}.

\bibitem{meltzoff1977imitation}
A.~N. Meltzoff and M.~K. Moore, ``Imitation of facial and manual gestures by human neonates,'' \emph{Science}, vol. 198, no. 4312, pp. 75--78, 1977.

\bibitem{davies2009investigating}
R.~Davies, E.~Kidd, and K.~Lander, ``Investigating the psycholinguistic correlates of speechreading in preschool age children,'' \emph{International Journal of Language \& Communication Disorders}, vol.~44, no.~2, pp. 164--174.

\bibitem{fu2019metricgan}
S.-W. Fu, C.-F. Liao, Y.~Tsao, and S.-D. Lin, ``Metricgan: Generative adversarial networks based black-box metric scores optimization for speech enhancement,'' in \emph{ICML 2019}, pp. 2031--2041.

\bibitem{prasad2021investigation}
A.~Prasad \emph{et~al.}, ``An investigation of end-to-end models for robust speech recognition,'' in \emph{IEEE ICASSP 2021}.

\bibitem{huang2013audio}
J.~Huang and B.~Kingsbury, ``Audio-visual deep learning for noise robust speech recognition,'' in \emph{IEEE ICASSP 2013}.

\bibitem{10127316}
A.~Das, S.~Patikar, and K.~Medhi, ``A survey on audio-visual speech recognition system,'' in \emph{2023 IEEE I3CS}, 2023.

\bibitem{wirth2022automatic}
J.~Wirth and R.~Peinl, ``Automatic speech recognition in german: A detailed error analysis,'' in \emph{2022 IEEE COINS}.

\bibitem{leng2021fastcorrect}
Y.~Leng \emph{et~al.}, ``Fastcorrect: Fast error correction with edit alignment for automatic speech recognition,'' \emph{NeurIPS 2021}.

\bibitem{mani2020asr}
A.~Mani, S.~Palaskar, N.~V. Meripo, S.~Konam, and F.~Metze, ``Asr error correction and domain adaptation using machine translation,'' in \emph{IEEE ICASSP 2020}.

\bibitem{chen2022wavlm}
S.~Chen \emph{et~al.}, ``Wavlm: Large-scale self-supervised pre-training for full stack speech processing,'' \emph{IEEE JSTSP 2022}.

\bibitem{ma2023auto}
P.~Ma \emph{et~al.}, ``Auto-avsr: Audio-visual speech recognition with automatic labels,'' in \emph{IEEE ICASSP 2023}.

\bibitem{watanabe2018espnet}
S.~Watanabe \emph{et~al.}, ``Espnet: End-to-end speech processing toolkit,'' \emph{arXiv preprint arXiv:1804.00015}, 2018.

\bibitem{librilight}
\BIBentryALTinterwordspacing
J.~Kahn \emph{et~al.}, ``Libri-light: A benchmark for asr with limited or no supervision,'' in \emph{IEEE ICASSP 2020}.\hskip 1em plus 0.5em minus 0.4em\relax IEEE, 2020. [Online]. Available: \url{http://dx.doi.org/10.1109/ICASSP40776.2020.9052942}
\BIBentrySTDinterwordspacing

\bibitem{wang2021voxpopuli}
C.~Wang \emph{et~al.}, ``Voxpopuli: A large-scale multilingual speech corpus for representation learning, semi-supervised learning and interpretation,'' \emph{arXiv preprint arXiv:2101.00390}, 2021.

\bibitem{gulati2020conformer}
A.~Gulati, J.~Qin, C.-C. Chiu, N.~Parmar, Y.~Zhang, J.~Yu, W.~Han, S.~Wang, Z.~Zhang, Y.~Wu \emph{et~al.}, ``Conformer: Convolution-augmented transformer for speech recognition,'' \emph{arXiv preprint arXiv:2005.08100}, 2020.

\bibitem{he2016deep}
K.~He, X.~Zhang, S.~Ren, and J.~Sun, ``Deep residual learning for image recognition,'' in \emph{IEEE CVPR 2016}.

\bibitem{Kim2016JointCB}
S.~Kim \emph{et~al.}, ``Joint ctc-attention based end-to-end speech recognition using multi-task learning,'' \emph{IEEE ICASSP 2017}.

\bibitem{son2017lip}
S.~et~al., ``Son chung, joon and others,'' in \emph{CVPR 2017}.

\bibitem{chung2018voxceleb2}
J.~S. Chung, A.~Nagrani, and A.~Zisserman, ``Voxceleb2: Deep speaker recognition,'' \emph{arXiv preprint arXiv:1806.05622}, 2018.

\bibitem{ephrat2018looking}
A.~Ephrat \emph{et~al.}, ``Looking to listen at the cocktail party: A speaker-independent audio-visual model for speech separation,'' \emph{arXiv preprint arXiv:1804.03619}, 2018.

\bibitem{yang2022audiovisual}
K.~Yang, D.~Markovic, S.~Krenn, V.~Agrawal, and A.~Richard, ``Audio-visual speech codecs: Rethinking audio-visual speech enhancement by re-synthesis,'' in \emph{CVPR 2022}.

\bibitem{donley2021easycom}
D.~et~al., ``Easycom: An augmented reality dataset to support algorithms for easy communication in noisy environments,'' 2021.

\bibitem{doi:10.1073/pnas.1612524113}
J.~Traer and J.~H. McDermott, ``Statistics of natural reverberation enable perceptual separation of sound and space,'' \emph{Proceedings of the National Academy of Sciences}, vol. 113, no.~48, pp. E7856--E7865, 2016.

\bibitem{gemmeke2017audio}
J.~F. Gemmeke \emph{et~al.}, ``Audio set: An ontology and human-labeled dataset for audio events,'' in \emph{IEEE ICASSP 2017}.

\bibitem{martinez2020lipreading}
B.~Martinez, P.~Ma, S.~Petridis, and M.~Pantic, ``Lipreading using temporal convolutional networks,'' in \emph{IEEE ICASSP 2020}.

\bibitem{ma2022visual}
P.~Ma \emph{et~al.}, ``Visual speech recognition for multiple languages in the wild,'' \emph{Nature Machine Intelligence}, vol.~4, no.~11, pp. 930--939, 2022.

\bibitem{ma2018shufflenet}
N.~Ma \emph{et~al.}, ``Shufflenet v2: Practical guidelines for efficient cnn architecture design,'' in \emph{ECCV 2018}.

\bibitem{zhao2023survey}
W.~X. Zhao \emph{et~al.}, ``A survey of large language models,'' \emph{arXiv preprint arXiv:2303.18223}, 2023.

\bibitem{xu2023parameter}
L.~Xu, H.~Xie, S.-Z.~J. Qin, X.~Tao, and F.~L. Wang, ``Parameter-efficient fine-tuning methods for pretrained language models: A critical review and assessment,'' \emph{arXiv preprint arXiv:2312.12148}, 2023.

\bibitem{zhang2024llamaadapter}
R.~et~al., ``{LL}a{MA}-adapter: Efficient fine-tuning of large language models with zero-initialized attention,'' in \emph{ICLR 2024}.

\bibitem{zhang2024tinyllama}
P.~Zhang, G.~Zeng, T.~Wang, and W.~Lu, ``Tinyllama: An open-source small language model,'' 2024.

\bibitem{textbooks2}
Y.~Li \emph{et~al.}, ``Textbooks are all you need ii: \textbf{phi-1.5} technical report,'' \emph{arXiv preprint arXiv:2309.05463}, 2023.

\bibitem{su2020hifi}
J.~Su, Z.~Jin, and A.~Finkelstein, ``Hifi-gan: High-fidelity denoising and dereverberation based on speech deep features in adversarial networks,'' \emph{arXiv preprint arXiv:2006.05694}, 2020.

\bibitem{afouras2018deep}
T.~Afouras, J.~S. Chung, A.~Senior, O.~Vinyals, and A.~Zisserman, ``Deep audio-visual speech recognition,'' \emph{IEEE TAPMI 2018}.

\bibitem{pan-etal-2022-leveraging}
X.~Pan \emph{et~al.}, ``Leveraging unimodal self-supervised learning for multimodal audio-visual speech recognition,'' in \emph{ACL 2022}, S.~Muresan, P.~Nakov, and A.~Villavicencio, Eds.

\bibitem{xu2020discriminative}
B.~Xu \emph{et~al.}, ``Discriminative multi-modality speech recognition,'' in \emph{Proceedings of the IEEE/CVF conference on Computer Vision and Pattern Recognition}, 2020, pp. 14\,433--14\,442.

\end{thebibliography}

\end{document}